\documentclass[aps,prd,reprint,superscriptaddress,showpacs,nofootinbib,longbibliography]{revtex4-1}
\usepackage{graphicx}
\usepackage{amsfonts}
\usepackage{amsmath}
\usepackage{mathrsfs}
\usepackage{epsfig}
\usepackage{slashed}
\usepackage{dcolumn}%

\usepackage{ulem}
\usepackage{color}
\usepackage{caption}
\usepackage{subcaption}
\usepackage{hyperref}

\newcommand\diff{\,\mathrm{d}}

\def\sign{\sigma_{\chi,n}}
\def\sigdm{\sigma_{\chi,\chi}}

\def\sige{\sigma_{\chi,e}}

\def\sigv{\langle\sigma_{\chi,\chi} v\rangle}

\def\nuflub8{\phi^\nu_B}
\def\nuflube7{\phi^\nu_{Be}}

\def\rhodm{\rho_{\chi}}
\def\rdm{r_{\chi}}
\def\mdm{m_{\chi}}
\def\ldm{l_{\chi}}
\def\lesim{\lesssim}
\def\grsim{\gtrsim}
\def\vstar{v_{*}}
\def\Vstar{V_{*}}
\def\vmean{\bar{v}}
\def\vesc{v_{\mathrm{esc}}}

\def\Rstar{R_{*}}
\def\Lstar{L_{*}}
\def\rhostar{\rho_{*}}
\def\Mstar{M_{*}}

\def\ndm{n_{\chi}}

\def\Ldm{L_{\chi}}
\def\Lscm{L_{\mathrm{SCM}}}
\def\dLscm{\dot{L}_{\mathrm{SCM}}}
\def\Lobs{L_{\mathrm{obs}}}
\def\dLobs{\dot{L}_{\mathrm{obs}}}
\def\Pobs{P_{\mathrm{obs}}}
\def\dPobs{\dot{P}_{\mathrm{obs}}}
\def\Pscm{P_{\mathrm{SCM}}}
\def\dPscm{\dot{P}_{\mathrm{SCM}}}
\def\Msun{M_{\bigodot}}

\def\Lsun{L_{\bigodot}}

\def\nat{Nature\ }
\def\aap{Astron.\ Astrophys.\ }
\def\aapr{Astron.\ Astrophys.\ Rev.\ }
\def\apj{Astrophys.\ J.\ }
\def\apjl{Astrophys.\ J.\ Lett.\ }

\def\aj{Astron.\ J.\ }
\def\mnras{Mon.\ Not.\ Roy.\ Astron.\ Soc.\ }
\def\physrep{Phys.\ Rept.\ }
\def\prd{Phys.\ Rev.\ D\ }

\def\araa{Annu.\ Rev.\ Astron.\ Astrophys.\ }
\def\jcap{J.\ Cosmol.\ Astropart.\ Phys.\ }

\def\pasp{Publications\ of\ the\ Astronomical\ Society\ of\ the\ Pacific}

\def\km{\,\mathrm{km}}
\def\TeV{\,\mathrm{TeV}}
\def\GeV{\,\mathrm{GeV}}
\def\MeV{\,\mathrm{MeV}}

\def\MV\TeV{\,\mathrm{MV}}
\def\cm{\,\mathrm{cm}}
\def\s{\,\mathrm{s}}

\newcolumntype{p}{D{,}{\pm}{-1}}

\def\c{\,\mathrm{c}}
\def\erg{\,\mathrm{erg}}
\def\g{\,\mathrm{g}}

\begin{document}

% The following information is for internal review, please remove them for submission
%\widetext
%\leftline{Version xx as of \today}
%\leftline{Primary authors: Joe E. Physics}
%\leftline{To be submitted to (PRL, PRD-RC, PRD, PLB; choose one.)}
%\leftline{Comment to {\tt d0-run2eb-nnn@fnal.gov} by xxx, yyy}
%\centerline{\em D\O\ INTERNAL DOCUMENT -- NOT FOR PUBLIC DISTRIBUTION}

% the following line is for submission, including submission to the arXiv!!
%\hspace{5.2in} \mbox{Fermilab-Pub-04/xxx-E}

\title{Probing the Dark Matter-Electron Interactions via Hydrogen-Atmosphere Pulsating White Dwarfs}

\author{Jia-Shu Niu}
\email{jsniu@sxu.edu.cn}
\affiliation{Institute of Theoretical Physics, Shanxi University, Taiyuan, 030006, China}
\affiliation{CAS Key Laboratory of Theoretical Physics, Institute of Theoretical Physics, Chinese Academy of Sciences, Beijing, 100190, China}
\affiliation{School of Physical Sciences, University of Chinese Academy of Sciences, No.19A Yuquan Road, Beijing 100049, China}
\author{Tianjun Li}
\email{tli@itp.ac.cn}
\affiliation{CAS Key Laboratory of Theoretical Physics, Institute of Theoretical Physics, Chinese Academy of Sciences, Beijing, 100190, China}
\affiliation{School of Physical Sciences, University of Chinese Academy of Sciences, No.19A Yuquan Road, Beijing 100049, China}
\author{Weikai Zong}
\affiliation{Department of Astronomy, Beijing Normal University,    Beijing 100875, China}
\author{Hui-Fang Xue}
\affiliation{Department of Astronomy, Beijing Normal University,    Beijing 100875, China}
\author{Yang Wang}
\affiliation{School of Mathematical Sciences, Shanxi University, Taiyuan 030006, China.}

%\ author_list.tex       % D0 authors (remove the first 3 lines
                             % of this file prior to submission, they
                             % contain a time stamp for the authorlist)
                             % (includes institutions and visitors)
\date{\today}

\begin{abstract}
In this work, we propose a novel scenario to probe the interactions between dark matter (DM) particles and electrons, via hydrogen-atmosphere pulsating white dwarfs (DAVs) in globular clusters. In this special configuration, the DM particles, which are predominantly captured by multiple scattering with the electrons in a DAV, would annihilate by pairs and provide extra energy source to the DAV. This mechanism slows down the natural cooling evolution which can be presented by the period variation rates of pulsation modes. The differences between the secular rates predicted by the precise asteroseismology and the secular rates obtained from observation can reveal the DM-electron interactions. An important observable has been proposed and corresponding estimations have been made. According to the estimation, if this scenario could be implemented in the near future, the potential sensitivity on $\mdm$ (DM particle's mass) and $\sige$ (elastic scattering cross section between DM and electron) could be hopefully extended to a region $5 \GeV \lesim \mdm \lesim 10^{4} \GeV$ and $\sige \grsim 10^{-40} \cm^{2}$. Combining  with indirect DM detection results, this could give us a cross check on the existence of such leptonphilic DM particles to some extent.
\end{abstract}

%\pacs{}
\maketitle

\section{Introduction}
Although Dark Matter (DM) contributes to $26.8 \%$ of the total energy density of the Universe \citep{Plank2014}, the particle nature of DM remains largely unknown. In recent years, the DM particle candidates have been searched for  via three main strategies, i.e., direct detection, indirect detection, and collider searches (see, e.g., \citep{Bertone2005} for reviews). For direct detection and collider searches, null results have been obtained yet, and the most stringent constraint from the current experiments on elastic scattering cross section between DM and nucleon is $\sign \lesim 4.1 \times 10^{-47} \cm^{2}$ for a DM particle's mass $\mdm = 30 \GeV$, and that between DM and electron is $\sige < 3 \times 10^{-38} \cm^{2}$ for $\mdm = 100 \MeV $ and $\sige < 10^{-37} \cm^{2}$ for $20 \MeV \le \mdm \le 1 \GeV$ \citep{XENON1T2018,XENON2012}.

An alternative DM search  can be carried out on the celestial objects, such as stars, which have huge volumes and large masses compared to the manual facilities. There are some branches which have been worked on:  the Sun and main-sequence stars, considered the precise properties of their interior structures from helioseismology and asteroseismology (see, e.g., \citep{Frandsen2010,Iocco2012,Vincent2015}); and  compact stars (i.e., white dwarfs (WD) and neutron stars (NS)), considered for their deficiency of nuclear energy source (see, e.g., \citep{Moskalenko2007,Bertone2008,Hurst2015,Amaro2016}). Nevertheless, most these attempts concentrate on the interactions between DM and nucleons whose cross section ($\sign$) has been constrained below $4.1 \times 10^{-47} \cm^{2}$ by direct detection \citep{XENON1T2018}. In order to enhance the interacting effects between DM and nucleons over the entire stars, a circumstance with high DM local density is preferred, such as the stars in  galaxy centers, globular clusters and dwarf galaxies.

The different schemes in this field are listed in Table \ref{tab:schemes}. In main sequence stars, the extra energy source from captured and then annihilated DM particles can be ignorable compared to the nuclear fusion, so one has to use the other DM particle models in which they cannot annihilate by pairs. As a result, in Scheme I we focus on the DM particles that cannot annihilate (i.e., asymmetric DM (ADM)), and consider the energy transferring effects of DM particles which lead to a change in the core structure (which can be detected by asteroseismology) of the star. In Scheme II, the absolute luminosities of the stars cannot directly detected in real cases, which is caused by the unknown distance to the star and the effects on dispersion. More importantly, we cannot determine the structure and constituents of the star only from the luminosity, which is the key point for estimating the DM captured effects. Consequently, in this Scheme, it is always suitable to do statistical research rather than individual ones. In Scheme III, the motivation is to explain the too fast cooling process of some pulsating WDs. The generation of axion in such stars increase the cooling rates of them, which in reverse can give constraints on axion mass from observations.

\begin{table*}[htb]
\begin{center}
\begin{tabular}{cccccccc}
  \hline\hline
Schemes  &DM models  &Techniques  &Stars  &Environments  &DM Effects  &Observables  &References  \\
  \hline
I   &ADM  &Asteroseismology  &Main sequence stars  &Usual  &Transferring energy  &$\Pobs$  &\citep{Frandsen2010,Iocco2012,Vincent2015}  \\
II   &WIMPs  &Direct observation  &Compact stars  &DM-dense  &Extra source  &$\Lobs$  &\citep{Moskalenko2007,Bertone2008,Hurst2015,Amaro2016}  \\
III   &Axions  &Asteroseismology  &Pulsating WDs  &Usual  &Bring energy  &$\Pobs$, $\dPobs$  &\citep{Corsico2012,Corsico2012b}  \\
This work   &WIMPs  &Asteroseismology  &DAVs  &Globular clusters  &Extra source  &$\Pobs$, $\dPobs$  &--  \\
  
  \hline\hline
\end{tabular}
\end{center}
\caption{The comparison between different Schemes in this field.}
\label{tab:schemes}
\end{table*}

White dwarfs (WDs) are thought to be the final evolutionary state of stars whose masses are not high enough to become  neutron stars or  black holes, which would include over 97$\%$ of the stars in the Milky Way \citep{Fontaine2001} and each of them is composed by an electron-degenerate core and an atmosphere envelope. They are considered to be the most electron-dense objects and can be the most promising laboratories to measure the DM-electron interactions.

The hydrogen-atmosphere pulsating white dwarfs (DAVs) are a type of pulsating WD with hydrogen-dominated atmospheres and the spectral type DA. Precise asteroseismology on DAVs can reveal their interior structures and determine the rates of the period variations which are related with their pulsation modes spanning over long time scales (see, e.g., \citep{Winget2008} for reviews). These secular rates reveal the evolutionary cooling rates of DAVs which can be described by the Standard Cooling Model (SCM). However, the DM particles, which are captured by multiple scattering with the DAVs' constituents, would annihilate by pairs and provide extra energy source to the stars. The natural cooling process would be slowed down and could be detected by measuring the secular rates of the period variations. In reverse, we can use these secular rates to constrain the interactions between DM and DAVs' constituents (nucleons and electrons).

In this  work, we  consider the DAVs in the central region of globular clusters and focus on the DM-electron interactions. The paper is organized as follows. We first introduce the capture rate of DM particles in DAVs and the rate of period variations of DAVs in Sec. II and Sec. III, respectively. Then we present the estimations and some discussions in Sec. IV and Sec. V. Finally, future utility of this scheme is given in Sec. VI.

\section{The Capture Rate of DM Particles in DAVs}

Galactic DM particles are inevitably streaming through any celestial object. Those particles will loose energy when they scatter with nucleons (which we mostly did not consider in this paper) or electrons inside the celestial object, leading to their speed decreasing. If the velocity of the DM particles reaches below the escape velocity, they will be ``captured,'' i.e., they become bound to the star. Regardless of the effect of evaporation which is not important in this paper where we consider the DM particle mass $m_{\chi} > 5 \GeV$ \citep{Gould1987b,Gould1990a}, the evolution of the total number of DM particles, $N_{\chi}$, inside the star (or any celestial object) can be written as
\begin{equation}
\dot{N}_{\chi}=\Gamma_{c}- 2 \Gamma_{a},
\label{eq_capture}
\end{equation}
where $\Gamma_{c}$ is the particle capture rate, $\Gamma_{a}=\frac{1}{2} C_{a} N_{\chi}^2$ is the annihilation rate in the total star, and  $C_{a}$ is the annihilation rate per pair DM particles. Therefore, we have $N_{\chi} = \Gamma_{c} \tau \tanh(\frac{t}{\tau})$
with the equilibrium time scale $\tau = \sqrt{\frac{1}{C_{a} \Gamma_{c}}}$. When the dynamic equilibrium state is reached, the DM capture rate is balanced by the annihilation one \citep{Griest1987}, i.e., $\Gamma_{c} = 2 \Gamma_{a}$.

In our case, (i) the mean free path of the DM particles in the star is small compared to the size of the star; (ii) DM particle mass is very large compared to the interaction constituent in the star (electrons). Consequently, we use the results recently developed by \citet{Bramante2017} to calculate the DM capture rate of the star, in which case the DM particles predominantly captured by scattering multiple times rather than only once. The DM capture rate obtained by N times scatterings ($\Gamma_{c}^{N}$) can be described by Eq. (22) in Ref. \citep{Bramante2017}, and the DM capture rate is the sum over all 
$N$ of the individual $\Gamma_{c}^{N}$, $\Gamma_{c} = \sum_{N=1}^{\infty} \Gamma_{c}^{N}$.

Further, we make some additional simplifications as well:  (i) a uniform distribution of matter in a DAV: $\rhostar (r) = \rhostar = \Mstar/\Vstar$ ($\rhostar$, $\Mstar$ and $\Vstar$ are the density, mass, and volume of a DAV, respectively); (ii) The same chemical composition over the entire scattering volume $\Vstar$; (iii) A uniform temperature profile calculated from Eq.~(\ref{eq:evelope_structure}) in DAV because of the extremely high thermal conductivity of an electron degenerate core; (iv) As DAVs are always electrically neutral, we use the values $\frac{1}{2} \frac{\Mstar}{m_{p}}$ and $\frac{1}{2} \frac{\rhostar}{m_{p}}$ as the total number of electrons ($N_{e}$)  and local number density of electrons ($n_{e}$) in a DAV, respectively ($m_{p}$ and $m_{e}$ are the mass of proton and electron). 

The Knudsen number $K$, which indicates the ``localization'' of the DM transport, is
\begin{equation}
  \label{eq:Knumber}
  K = \frac{\ldm (0)}{\rdm},
\end{equation}
where $\ldm (0)=\left[ \sige \cdot n_{e}(0) \right]^{-1} = \left[ \sige \cdot n_{e}\right]^{-1}$ is the mean free length in the center of the star and $\rdm = \sqrt{\frac{3 k T_{c}}{2 \pi G \rho_{c} \mdm}}$ is the typical scale of the DM core in the star. Here $T_{c}$ and $\rho_{c}$ are respectively the temperature and density of the star's core, $G$ is the gravitational constant and $k$ is the Boltzmann constant.

Following Refs. \citep{Griest1987,Scott2009,Taoso2010}, in the case of large $K$ (for DAVs), the DM particles' distribution in star can be described by
\begin{equation}
\ndm (r)=\ndm (0) \cdot \exp \left[ - (\frac{r}{\rdm})^2  \right].
\label{eq:dis_K}
\end{equation}

The annihilation term can be computed by a separate way as follows:
\begin{equation}
\Gamma_{a} = \int_{0}^{\Rstar}  \diff r \ 4 \pi r^{2} \cdot \frac{1}{2} \sigv \ndm^{2} (r).
\label{eq:ann}
\end{equation}
The factor 1/2 (1/4) in the equation above is appropriate for self (nonself) conjugate particles
and $\sigv$ is the velocity-averaged DM annihilation cross section ($\sigdm$) multiplied by DM relative velocity ($v$).

If an equilibrium between capture and annihilation is reached the annihilation rate reduces to $\Gamma_{a} = 1/2 \Gamma_{c} $ and it is independent on the annihilation cross section. With the value of $\sigv \simeq 3 \times 10^{-26} \ \cm^{3} \s^{-1}$, we can impose  $\Gamma_{a} = 1/2 \Gamma_{c} $  to do the normalization and get $\ndm(0)$. 
Thus, the distribution $\ndm (r)$ is specified and all the related values in this equilibrium state are known.

\section{The Rate of Period Variations of DAVs}

The period variation of a DAV is related to two physical processes
in the star: the cooling of the star and the contraction of its atmosphere, and is given by
\begin{equation}
  \frac{\dot P}{P} \simeq -a \frac{\dot T_{c}}{T_{c}} + b \frac{\dot R}{R},
  \label{eq:var_period}
\end{equation}
where $P$ is the pulsation period for the $m=0$ multiplet component, $T_{c}$ is the maximum (normally, core) temperature, $R$ is the stellar radius, and $\dot P$, $\dot T$ and $\dot R$ are the respective temporal variation rates \citep{Winget2008}. The constants $a$ and $b$ are positive numbers of order unity. For DAVs, cooling dominates over gravitational  contraction, in such a way that the second term in Eq.~(\ref{eq:var_period}) is usually negligible, and only positive values of the observed period variation rate are expected \citep{Winget2008,Fontaine2008,Althaus2010,Calcaferro2017}.

From the structure of a WD's envelope, we have \citep{Kippenhahn1990}
\begin{equation}
  \label{eq:evelope_structure}
  T_{0}^{\frac{7}{2}} = B \frac{\Lstar/\Lsun}{\Mstar/\Msun},
\end{equation}
where $B \simeq 1.67 \times 10^{27}$ is a constant and $T_{0}$ is the interface temperature between the core and envelope, 
and  $\Lstar$ and $\Lsun$ are the luminosities of the star and sun, respectively.

If we  use the approximation $T_{0} \simeq T_{c}$ (for DAVs) in Eq.~(\ref{eq:evelope_structure}),  
substitute the result into Eq.~(\ref{eq:var_period}) and ignore the mass variation term during the cooling \citep{Winget2008,Fontaine2008,Althaus2010}, we obtain
\begin{equation}
  \label{eq:p_l}
  \frac{\dot P}{P} \simeq - \frac{2a}{7} \frac{\dot{L}_{*}}{\Lstar}.
\end{equation}

According to the annihilation of DM in a DAV, if the equilibrium state has been reached, the luminosity of the DAV should be $\Lobs = \Lscm + \Ldm$, where $\Lobs$ is the total observed luminosity of the DAV, $\Lscm$ is the normal luminosity in the SCM, and $\Ldm$ is the luminosity purely due to the annihilation of DM.

From the above section, we get $\Ldm = 2 \cdot \Gamma_{a} \mdm \c^{2} = \Gamma_{c} \mdm \c^{2}$.
One should note that once the equilibrium state is reached, $\Ldm$ should not change with time. 
As a result, we have $\dLobs = \dLscm$. Thus, the relationship between the rate of period variation 
and the luminosity should be (replace $\Lstar$ with $\Lobs$)
\begin{equation}
  \label{eq:L_P_result}
  \frac{\dPobs}{\dPscm} = \frac{\Lscm}{\Lscm + \Ldm},
\end{equation}
in which we impose that the period from model calculation $\Pscm$ equals to the value from observation $\Pobs$. 
Here, we ignore the effects of DM on the period of pulsation, because (i) the total mass of the DM particles in such DAV can be ignored compared to the star mass (see below for the estimation), and we can ignore their gravitational effects on star's pulsation; (ii) the period of a star is a dynamical quantity which is determined by its interior structure, and is not directly related to the energy injection in the star.

In SCM, the period variation of a DAV depends on the stellar mass and core composition, and can be expressed as a function of the mean atomic weight $A$ \citep{Kawaler1986, Kepler1995}
\begin{equation}
  \label{eq:dpdt}
  \dPscm = \diff P / \diff t = (3-4) \times 10^{-15} \frac{A}{14}\  \s \ \s^{-1}.
\end{equation}
in this work, we use a mean atomic weight of 14, which is consistent with the cooling rate of DAVs with a carbon-oxygen core \citep{Mukadam2013}.

\section{Estimation}

In this work, we consider the DAVs in the central region of globular clusters and focus on the DM-electron interactions. Globular clusters are always considered as the local DM-dense environments and the velocities between their member stars and the DM subhalo surrounding them can be measured precisely \citep{Merritt1997,Van2006}. Both of them can increase the DM capture rate of DAVs.

Although we have no identified DAVs in globular clusters yet due to the lack of scientific aims to do long time-series photometry observations by large aperture telescopes, we can put a well-studied DAV in a well-studied globular cluster to do the estimation.

G117-B15A, whose $\dPobs$ matches $\dPscm$ well within uncertainties \citep{Chen2017} and its structure details have been determined by asteroseismology in Refs. \citep{Romero2012,Corsico2012,Chen2017}. Here we use the structure details from Table 5, Ref. \citep{Chen2017}.

In order to get large DM density, we choose $\omega$ Cen as the globular cluster to do the estimation. \citet{Amaro2016} has estimated $\rhodm \simeq 4 \times 10^{3} \GeV \cm^{-3} $ near the center of $\omega$ Cen without an intermediate-mass black hole (IMBH) and $4 \times 10^{3} \GeV \cm^{-3} \lesim \rhodm \lesim 4 \times 10^{9} \GeV \cm^{-3}$ with an IMBH. We here choose $\rhodm = 4 \times 10^{3} \GeV \cm^{-3} $ to do the estimation. From Refs. \citep{Merritt1997,Van2006}, the members of this cluster are orbiting the center of mass with a peak velocity dispersion  $\vstar \simeq 7.9 \km \s^{-1}$. Near a IMBH, where orbital motion around a single mass dominates, the test particle (DM or star) velocities are Keplerian, $\vstar = \vmean$ ($\vmean$ is the mean velocity of the test particles).

 With above configurations, if we fix the $\mdm = 100 \GeV$ and $\sige = 10^{-38} \cm^{2}$, we obtain the luminosity with DM annihilation of the DAV as $\Ldm \simeq 1.04 \times 10^{31} \erg \s^{-1}$, which is about one order larger than the $\Lscm \simeq 1.23 \times 10^{30} \erg \s^{-1}$. As a result, the period variation of DAV should be $\dPobs \simeq 0.1 \times \dPscm \simeq (0.3-0.4) \times 10^{-15} \frac{A}{14} \s \s^{-1} \simeq (0.3-0.4) \times 10^{-15} \s \s^{-1}$, which is obviously smaller than the value from SCM \citep{Winget2008} (see  Eq.~(\ref{eq:dpdt})).

 Moreover, we get the equilibrium time scale $\tau \simeq 4.89\ \mathrm{yr}$, which is really a short period compared with the time scale of  DAV formation processes. Thus, we can consider DAVs to be always in the state of DM capture and annihilation equilibria which has been assumed in this work. The total DM mass in the DAV is about $1.8 \times 10^{18} \g \ll \Mstar $, and then  its gravitational effects on the DAV's interior structure can be neglected, which is consistent with our assumption $\Pscm = \Pobs$ to determine the interior structure of a DAV.

Furthermore, we make a similar estimation with the same configurations as above considering the DM-nucleon interactions. In this case, the formula in Ref. \citep{Gould1987a} was used, which is constructed with the single scattering  capture for the case that the mean free path of the DM particles in DAV is much larger compared to the size of the star. At last, we obtain $\Ldm \simeq 1.49 \times 10^{24} \erg \s^{-1}$ (here we choose $\sign = 10^{-46} \cm^{2}$ and $\mdm = 100 \GeV$), which can be ignored compared  with the $\Ldm$ from electrons. Consequently, it is a reasonable assumption to consider DM captured by interactions with electrons alone in a DAV.

\section{Discussions}
In this work, benefiting from (i) the large DM local density in the central region of globular clusters; (ii) The relatively small velocity dispersion of the members in globular clusters; (iii) The large escape velocity of DAVs; (iv) The large electron density in DAVs; (v) The DM particle  multiscattering captured by DAVs,
we found that the luminosity due to the DM annihilation would be so large ($\Ldm / \Lscm \sim 10$) in our estimation. 
But in real cases, the absolute luminosity of a star cannot be determined directly due to the unknown distance to the star and the effects on dispersion. More importantly, we cannot determine the structure and constituents of the star only from the absolute luminosity, which is the key point for estimating the DM captured effects. Consequently, some of the previous works \citep{Moskalenko2007,Bertone2008,Hurst2015,Amaro2016} which detect DM annihilation in compact stars according to directly observing their luminosities are difficult to achieve, and they are only suitable to do statistical researches rather than individual ones under the precondition of calibrating the stars' relative luminosity.

Fortunately, a DAV's internal structure and long time evolution in SCM (represented by $\dPscm$) can be well modeled by precise asteroseismology, which just need the data from high-precision time-series photometry. At the same time, these photometry data can also provide us $\dPobs$, which represents the real long time evolution of a DAV. Compared with these two long time scale quantities, the influence from the DM-electron interaction would be probed.

\begin{figure}
\centering
\includegraphics[width=0.48\textwidth]{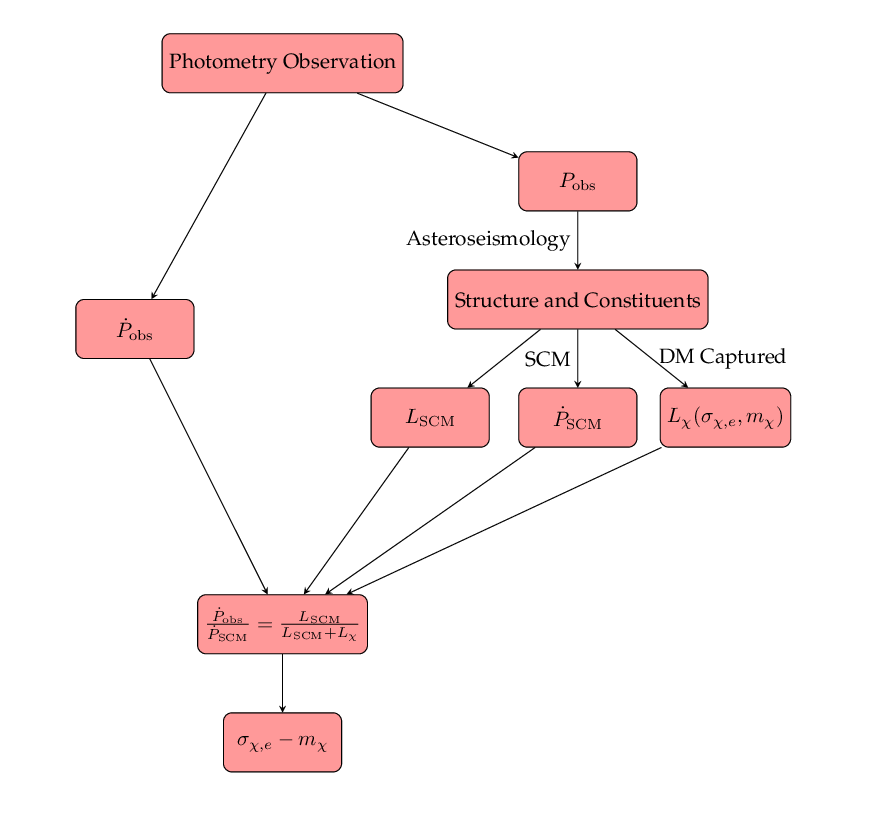}
\caption{The flow diagram of this work.}
\label{fig:flow_diagram}
\end{figure}

 The flow diagram is presented in Fig. \ref{fig:flow_diagram}. From Fig. \ref{fig:flow_diagram}, we can find the uncertainties of the final results from $\Lscm$, $\dPscm$, $\Ldm$ and $\dPobs$. In a classical asteroseismology on a DAV (see, e.g., \citep{Corsico2012}), the uncertainties for both $\Lscm$ and $\dPscm$ are $7\%$, and $\dPobs$ can be determined from observation within $17\%$ uncertainty. In a multiscattering captured process, the scattering times $N$ is always chosen to be the cutoff at a large number (in this paper, we choose $N=100$, which leads to an underestimate of $\Ldm$ by a factor $\lesim 0.001\%$), but this uncertainty can be suppressed to be less than any given small value theoretically. Another type of uncertainty comes from the physical properties of the globular cluster where the DAVs are located. Taking $\omega$ Cen as an example, if the distribution model of DM particles is given, we find that the DM local density $\rhodm$ can be gotten with an uncertainty of $\sim 20\%$ \citep{Souza2013}. The velocity dispersion $\vstar$ (also $\vmean$) has an uncertainty of $30\%$. Because $\vesc \gg \vstar$ in our case, its uncertainty has a negligible influence on $\Ldm$.  

Here, we want to emphasize that, because we have never done long time-series photometric observations on the DAVs in globular clusters, the relevant uncertainties of the DAVs in the above estimations are obtained from the local ones which have been studied successfully by asteroseismology based on long time-series observations. These observations are always implemented by ground-based optical telescopes of about 1-3 meters. We hope that the new generation ground-based or space optical telescopes would give us a comparable or even better result on the DAVs in globular clusters.

\subsection{Relations to DM Indirect Detection}

The observations from high-energy cosmic ray (CR) spectra provide us some more important hints. Although the antiproton spectrum~\citep{AMS02_pbar_proton} has shown some excess which cannot explained by conventional propagation models (see, e.g., \citep{Niu2017a}), and some works (see, e.g., \citep{Cuoco2017,Cui2017}) propose DM interpretations. But if we consider the uncertainties from the antiproton production cross section~\citep{Lin2016} and the hardening of the primary source injection spectra of proton and helium~\citep{Tomassetti2012,Tomassetti2015prd,Niu2018}, the excess in the antiproton spectrum cannot give us a concrete conclusion. At the same time, the high-energy electron/positron spectra show a obvious excess confirmed by many experiments (such as ATIC~\citep{ATIC_lepton}, PAMELA~\citep{PAMELA_lepton}, AMS-02~\citep{AMS02_lepton,AMS02_lepton_sum,AMS02_fraction01,AMS02_fraction02}, and DAMPE~\citep{DAMPE2017}), which is a big unsolved problem in CR physics. This anomaly can be interpreted by both astrophysical scenarios (such as pulsar and Supernova Remnants, see, e.g. \citep{Shen1970,Yuksel2009,Blasi2009}) and DM scenario (annihilation and decay, see, e.g. \citep{Bergstrom2008,Zhang2009,Bergstrom2009}). Although recent work from \citet{HAWC2017} claimed that local pulsars could not contribute enough positrons to reproduce the observed CR positron spectrum, the origin of the positron excess is still unclear.

  If we want to ascribe the electron/positron excess to DM annihilation, these DM particles should annihilate mainly (or totally) via lepton channels \footnote{Because if it is not so, there should be obvious excess in the CR antiproton spectrum.}. Such kinds of DM particles (always called leptonphilic DM particles, see for, e.g., \citep{Chen:2008dh,Yin:2008bs,Bi:2009md,Fox:2008kb,Cao:2009yy}) should interact mainly (or totally) with leptons other than nucleons. This means that, for such kinds of DM particles, the effective elastic scattering cross section between DM and leptons ($\sige$) should be much larger than that between DM and nucleons ($\sign$). Because the relation between $\sige$ and $\sigdm$ \footnote{The DM particle annihilation cross section, which can be derived from CR lepton spectra.} is based on specific DM particle models, the measurements on $\sige$ could not only give a cross-check on the existence of such leptonphilic DM particles, but also give us hints to construct relevant DM particle models as well.

Unfortunately, the DM particle's mass derived from lepton spectra is always in the range of about $100 \GeV - 10 \TeV$ ~\citep{Bergstrom2008,Zhang2009,Bergstrom2009,Lin2015,Niu2017b,Niu2017c}, which is out of the current direct detectors' sensitivity on DM-electron interactions ($\le 1 \GeV$). Consequently, new detecting scenarios should be implemented, and the scheme in this work could give a cross-check on the existence of such DM particles to some extent.

\section{Future Utility}

\subsection{Observing DAVs in Globular Clusters}
\label{sec:DAVinGC}

\begin{figure}
\centering
\includegraphics[width=0.49\textwidth]{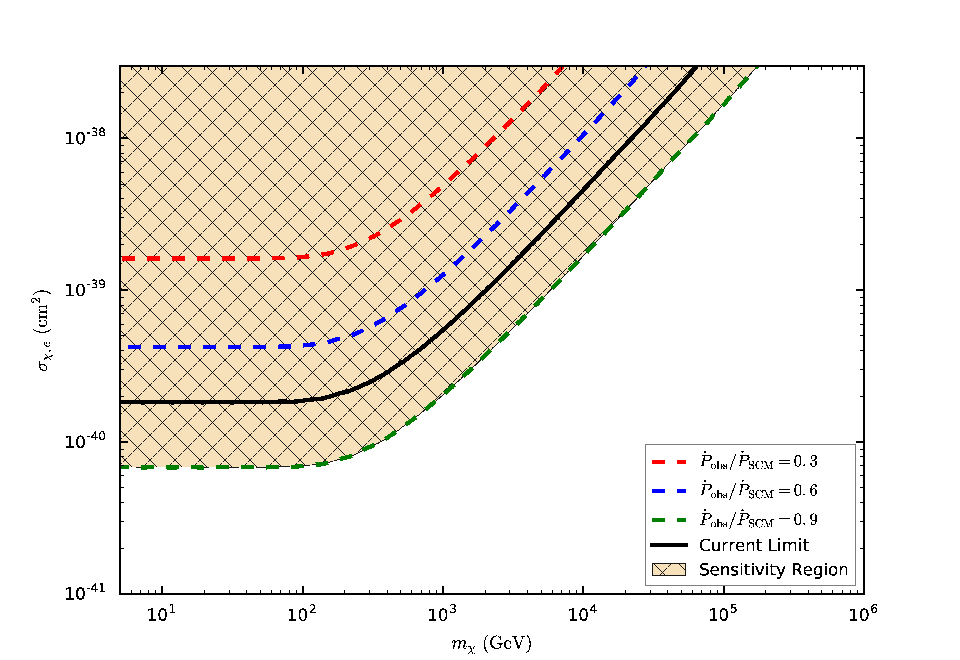}
\caption{ The prospective constraints on $\sige - \mdm$ from our estimations. 
The lower limit of the exclusive line is determined by the uncertainty from $\dPobs/\dPscm$. The black line represents the exclusive line based on current uncertainty\footnote{Here, "current uncertainty" means the uncertainty from the observation and model construction of G117-B15A, which represent a classical level on current asteroseismology of DAVs.} on $\dPobs/\dPscm$, and the colored dashed lines correspond to  specific $\dPobs/\dPscm$ values. In the near future, the exclusive line would be extended to the line $\dPobs/\dPscm = 0.9$.}
\label{fig:sigma_mass}
\end{figure}

  Considering all the uncertainties listed above, based on current observations, if we could find a DAV like G117-B15A in such DM-dense environment (in the central region of $\omega$ Cen), and do successful asteroseismology according to the observation, the prospective sensitivity region on $\mdm - \sige$ diagram can be gotten in Fig. \ref{fig:sigma_mass}. In this figure, each value of $\dPobs/\dPscm$ corresponds to a line on $\mdm - \sige$ diagram. Considering a value of $\dPobs/\dPscm$ with uncertainties from observations, we could obtain a band in  which the DM particles live.

  In the near future, based on the precise measurements of the distance, the proper motion and radial velocity by Gaia (which has already launched in 2013) and the accumulation of high-precision time-series photometry data from Gaia, the Transiting Exoplanet Survey Satellite (TESS, which has already launched in 2018) or the James Webb Space Telescope (JWST), the uncertainty of $\dPobs/\dPscm$ could hopefully achieve a level of $\lesssim 10 \%$. In such cases,  the detectable DM-electron interaction parameter space could be extended to $5 \GeV \lesim \mdm \lesim 10^{4} \GeV$ and $\sige \grsim 10^{-40} \cm^{2}$.

  If the observed result is positive ($\dPobs/\dPscm$ has a high confidential less than 0.9), the existence of leptonphilic DM particles could be confirmed to some extent, and the allowed parameter space on $\mdm - \sige$ could be determined. Furthermore, combining with the allowed parameter space from CR lepton spectra, we would determine the value of $\mdm$, $\sige$, $\sigdm$ and construct relevant DM particle models.

  On the other hand, if the observed result is negative ($\dPobs/\dPscm$ is approximately equal to or larger than 0.9), some of the leptonphilic DM-particle models could be excluded, just like current situations in DM direct detection.

\subsection{Observing Nearby DAVs }
\label{sec:DAVinlocal}

Recently released DAMPE lepton spectrum \citep{DAMPE2017} shows a tentative peak at $\sim 1.4 \TeV$ which attracts many works (see, e.g., \citep{Yuan2017,Ge2017,Huang2017,Fowlie2017} ) to interpret. As the statements listed above, both astrophysical and DM scenarios can give an explanation for the peak to some extent. Although the confidential level of this peak signal is about $2-3 \sigma$ \citep{Huang2017,Fowlie2017}, as the accumulation of the counts, it will give us clearer results. If the peak signal is proved to be true, local DM substructures are needed to perform explanations, whose DM local density of their central region can reach up to $2 \times 10^{3} \GeV \cm^{-3}$ \citep{Diemand2008}. If the widely spread DAVs are located in them, we could discover these nearby DM substructures according to the observation of these nearby DAVs. This would provide another independent method to discover and confirm nearby DM substructures. Additionally, this task can be implemented by many ground-based optical telescopes.

\section*{Acknowledgments}

This research was supported in part by the National Science Foundation of China Projects No. 11475238 and No. 11875062, Key Research Program of Frontier Sciences, Chinese Academy of Sciences No. 11747601. We would like to thank Marina Cerme{\~n}o for her helpful suggestions. The numerical results described in this paper have been obtained via the HPC Cluster of ITP-CAS.

%%%\bibliographystyle{prl}
%%%\bibliography{dm_astero}% Produces the bibliography via BibTeX.



\end{document}